\documentstyle[aps,pra,twocolumn,epsf]{revtex}
\begin{document}

\twocolumn
[ \hsize\textwidth\columnwidth\hsize
\csname@twocolumnfalse\endcsname

\title{Quantum versus classical descriptions
of\\ sub-Poissonian light generation in three-wave mixing}

\author{Ji\v r\'\i\ Bajer$^1$ and Adam Miranowicz$^{2,3}$}

\address{$^{1}$Department of Optics, Palack\'{y} University,
17.~listopadu 50, 772~00 Olomouc, Czech Republic}
\address{$^{2}$CREST Research Team for Interacting Carrier Electronics,
School of Advanced Sciences,\\ The Graduate University for
Advanced Studies (SOKEN), Hayama, Kanagawa 240-0193, Japan}
\address{$^{3}$Nonlinear Optics Division, Institute of Physics,
Adam Mickiewicz University, 61-614 Pozna\'n, Poland}

\date{\today}
\maketitle

\begin{abstract}
Sub-Poissonian light generation in the non-degenerate three-wave
mixing is studied numerically and analytically within quantum and
classical approaches. Husimi $Q$-functions and their classical
trajectory simulations are analysed to reveal a special regime
corresponding to the time-stable sub-Poissonian photocount
statistics of the sum-frequency mode. Conditions for observation
of this regime are discussed. Theoretical predictions of the Fano
factor and explanation of the extraordinary stabilization of the
sub-Poissonian photocount behavior are obtained analytically by
applying the classical trajectories. Scaling laws for the maximum
sub-Poissonian behavior are found. Noise suppression levels in
the non-degenerate vs degenerate three-wave mixing are discussed
on different time scales compared to the revival times. It is
shown that the non-degenerate conversion offers much better
stabilization of the suppressed noise in comparison to that of
degenerate process.

\vspace{5mm} {\bf Keywords}: sub-Poissonian statistics, three-wave
mixing, Fano factor, classical trajectories

\end{abstract}

 ]

\section{Introduction}

For almost four decades, since the pioneering experiments of
Franken et al. \cite{Franken} and theoretical foundations laid
down by Bloembergen et al.\cite{Bloembergen}, multiwave mixing has
unceasingly been in the forefront of quantum-optical
investigations \cite{Barakat,Boyd}. In particular, the {\em
three-wave mixing} (TWM) has attracted considerable interest as a
parametric non-linear process of conversion of two sub-frequency
(say, $\omega_1$ and $\omega_2$) photons into one sum-frequency
($\omega _{1}+\omega _{2}\rightarrow \omega _{3}$) photon,
together with the inverse process. TWM can be observed in
non-linear crystals like ADP, KDP, LiNbO$_{3}$ or BaTiO$_{3}$
\cite{Dmitriev}. Both the total energy, $\hbar \omega _{1}+\hbar
\omega _{2}=\hbar \omega _{3}$, and momentum, $\hbar {\bf
k}_{1}+\hbar {\bf k}_{2}=\hbar {\bf k}_{3}$, of interacting
photons are conserved. TWM is observable for proper orientations
of light beam polarizations and crystal axes \cite{Boyd},
therefore it can be considered as a parametric process. TWM is
used for the frequency-up conversion if $\omega _{1}\rightarrow
\omega _{3}$ or the frequency-down conversion if $\omega
_{3}\rightarrow \omega _{1}$. The process is also useful for
generation of non-classical light such as squeezed, sub-Poissonian
and antibunched light \cite{Mandel}.

Before the computer era, the quantum dynamics was usually
investigated under the short-time approximation only. Nowadays,
the Taylor series of quantum operators can be found for almost any
number of terms with the help of fast computers and sophisticated
software. However, these series are usually convergent for short
evolution times or even for initial time only. Thus, numerical
quantum methods (see, e.g., \cite{Bajer91}) fail in simulation of
the long-time quantum evolution. On the other hand, as we have
shown in \cite{Bajer99,Bajer00}, the method of classical
trajectories gives very good estimation in the case of
strong-field interaction (practically, for photon numbers larger
than 10). The computational speed of the classical-trajectory
method does not depend on the numbers of interacting photons and,
moreover, for a larger number of photons one obtains better
precision. Thus, the method is very fast and offers a simple
substitute for the tedious exact quantum numerical calculations.
The classical-trajectory method enables not only numerical but
also some analytical predictions, e.g., for stationary Fano
factors \cite{Bajer99,Bajer00} or for maximum pump depletion in
TWM \cite{Bandilla1}. A method similar to ours to simulate
classical noise in TWM was used by Chmela \cite{Chmela}.

In the previous papers, we have studied degenerate processes of
wave mixing, including the second \cite{Bajer99} and higher
\cite{Bajer00} harmonic generations. Here, we generalize the
former results for the non-degenerate wave mixing. It is
well-known that both degenerate and non-degenerate TWM can be used
for generation of sub-Poissonian light \cite{Perina,Kozierowski}.
Nevertheless, theoretical predictions of quantum parameters, like
Fano factor, are most often derived under the short-interaction
(short-time or short-length) approximation (see, e.g.,
\cite{Perina,Kozierowski,Bajer92}), thus valid for weak
non-linear coupling of the optical fields only. Motivated by
papers of Nikitin and Masalov \cite{Nikitin} and of Bandilla,
Drobn\'y and Jex \cite{Bandilla,Drobny}, we analyse the
long-interaction evolution of TWM. The main result of this
article can be summarized as follows: the TWM can be a source of
time-stable sub-Poissonian light of the sum-frequency mode in the
no-energy-transfer regime. The deepest noise reduction, with the
Fano factor equal to 5/6, can be observed for the balanced input
amplitudes $r_{1}=r_{2}=r_{3}/\sqrt{2}$.  The same degree of
photocount noise suppression in the sum-frequency mode can be
achieved for the degenerate TWM. However, the sub-Poissonian light
produced in non-degenerate TWM is much better stabilized compared
to that in degenerate TWM. Moreover, the Fano factors for the
sub-frequency modes in the non-degenerate TWM are smaller than
those for the degenerate process. This and other results will be
demonstrated analytically by applying a method of classical
trajectories and tested numerically within quantum approach.

\section{Quantum analysis}

In quantum approach, the non-degenerate three-wave mixing can be
described by the interaction Hamiltonian (e.g., \cite{Perina})
\begin{equation}
\hat{H}=\hbar g\left( \hat{a}_{1}\hat{a}_{2}\hat{a}_{3}^{\dag
}+\hat{a}_{1}^{\dag }\hat{a}_{2}^{\dag }\hat{a}_{3}\right) ,
\label{N01}
\end{equation}
where $\hat{a}_{k}$ and $\hat{a}^{\dag}_{k}$ denote, respectively,
annihilation and creation operators of the sub-frequency (labeled
with subscripts 1, 2) and sum-frequency (subscript 3) modes; $g$
is a non-linear coupling parameter, which is related to the
quadratic susceptibility tensor $\chi ^{\left( 2\right) }$ of a
given non-linear optical crystal and also dependent on the
geometry of laboratory set-up \cite{Boyd}.

As in \cite{Bajer00}, we analyse the quantum Fano factors given by
$F_k=(\langle \hat{n}_k^{2}\rangle -\langle \hat{n}_k\rangle
^{2})/\langle \hat{n}_k\rangle$ for a photon-number operator
$\hat{n}_k=\hat{a}_k^{\dag} \hat{a}_k$. The light is referred to
as sub-Poissonian if $F_k<1$ and super-Poissonian if $F_k>1$.

For weak non-linear interactions or short crystal, the short-time
approximation can be applied for analytical predictions of
photocount noise suppression with $F{}_{k}<1$
\cite{Perina,Kozierowski}. The Fano factors are approximated by
\begin{eqnarray} \label{N02}
F{}_{1,2} &=&1+2r_{3}^{2}\,(gt)^{2} +\frac{8}{3}
r_{1}r_{2}r_{3}\sin \theta\, (gt)^{3}+{\cal O}\{(gt)^4\},
\nonumber \\
F{}_{3} &=& 1-\frac{4}{3}r_{1} r_{2} r_{3}\sin
\theta\,(gt)^{3}+{\cal O}\{(gt)^4\},
\end{eqnarray}
where $r_{k}$ are the input coherent amplitudes and $\theta =\phi
_{1}+\phi _{2} -\phi _{3}$ is the input phase mismatch. For $\sin
\theta >0$, the sub-Poissonian statistics in the sum-frequency
mode can be observed. For $\theta =0$, we find the higher-order
short-time Fano factor expansion to be
\begin{eqnarray*}\label{N03}
F{}_{3}=1+ \left( r_{3}^{2}-7
r_{1}^{2}r_{2}^{2}+4r_{1}^{2}r_{3}^{2}+4r_{2}^{2}
r_{3}^{2}\right)\frac{(gt)^{4}}{3} +{\cal O}\{(gt)^5\!\}.
\end{eqnarray*}
It is seen that the sub-Poissonian light in the sum-frequency mode
is generated for some combinations of input amplitudes $r_{k}$.
Since the Fano factors depend weakly on time (i.e., in its third
or higher-order power), it is difficult to observe the
sub-Poissonian light generation in the short-time regime.

To analyse the exact quantum dynamics of the TWM process beyond
the short-time approximation, we apply%
\begin{figure}
\vspace*{-4mm} \hspace*{-3mm} \epsfxsize=8.5cm\epsfbox{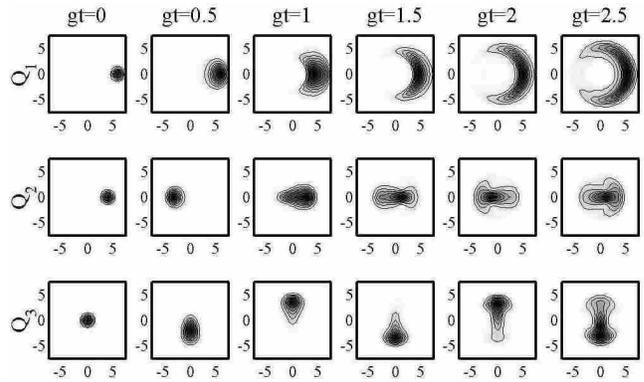}\vspace*{2mm}%
\caption{Quantum dynamics {\em out} of NETR: Cross-sections of
single-mode $Q$-functions: $Q_1({\rm Re}\, \alpha_1,{\rm Im}\,
\alpha_1)$ and $Q_2({\rm Re}\, \alpha_2,\allowbreak {\rm
Im}\,\alpha_2)$ for sub-frequency modes and $Q_3({\rm Re}\,
\alpha_3,{\rm Im}\,\alpha_3)$ for sum-fre\-quency mode at
different scaled evolution times for initial coherent fields with
real amplitudes, $\alpha_k(0) = r_k$, set to $r_{1}=6$, $r_{2}=4$
and $r_{3}=0$.}
\end{figure}
\begin{figure}
\vspace*{-5mm} \hspace*{-3mm} \epsfxsize=8.5cm\epsfbox{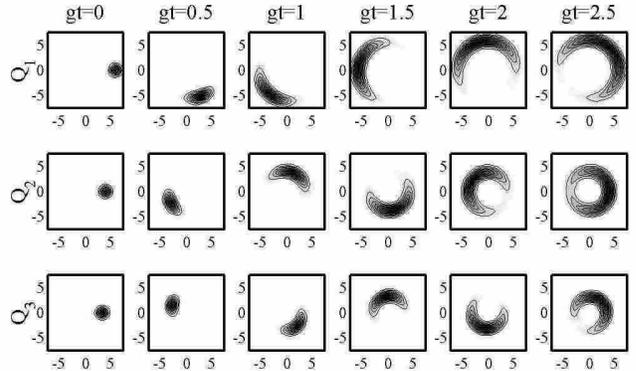}\vspace*{3mm}%
\caption{Quantum dynamics {\em in} NETR: Cross-sections of
marginal $Q$-functions as in figure 1, but for $r_{1}=6$,
$r_{2}=4$, and $r_{3}=r_1r_2/ \sqrt{r_1^2+r_2^2}\allowbreak\approx
3.328$. }
\end{figure}
\begin{figure}
\vspace*{-4mm} \hspace*{-7mm} \epsfxsize=9cm\epsfbox{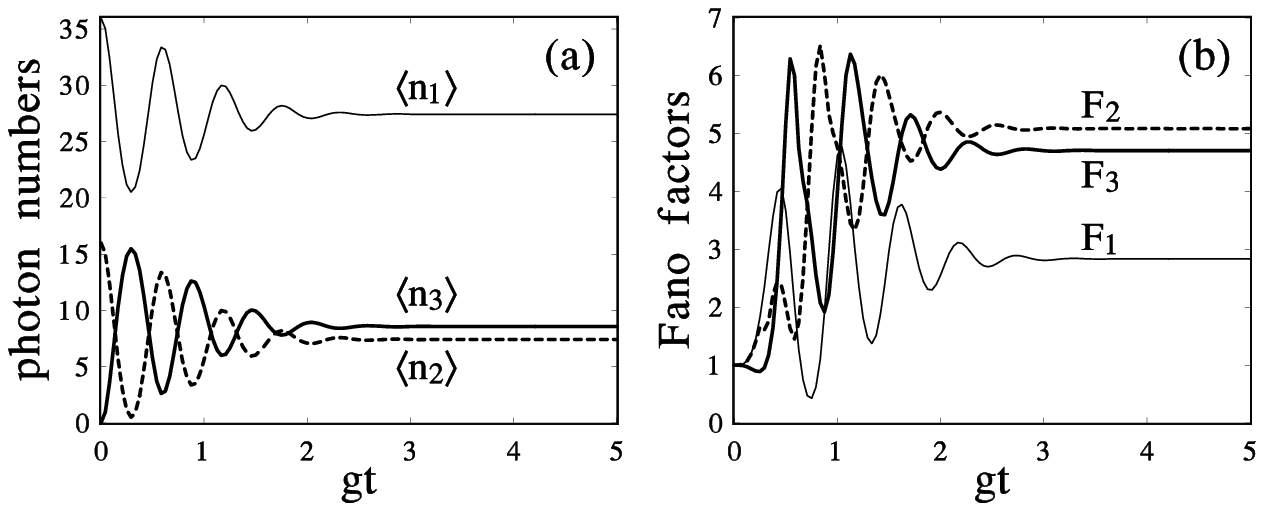}%
\caption{Quantum dynamics {\em out} of NETR: (a) photon
numbers $\langle\hat{n}_k\rangle$ and (b) Fano factors
$F_{k}$ ($k=1,2,3$) for coherent inputs with real amplitudes
$r_{1}=6$, $r_{2}=4$, and $r_{3}=0$. For longer times, all the
modes become stationary with the super-Poissonian statistics,
$F_{k}>1$.}
\end{figure}
the Walls-Barakat method \cite{Barakat} of Hamiltonian
diagonalization for the initial coherent states. Quantum analysis
enables numerical estimation of all statistical properties
including photocount noise. Complete quantum information of the
TWM dynamics can be given by the Husimi%
\begin{figure}
\vspace*{-3mm}
\hspace*{-4mm} \epsfxsize=8.7cm\epsfbox{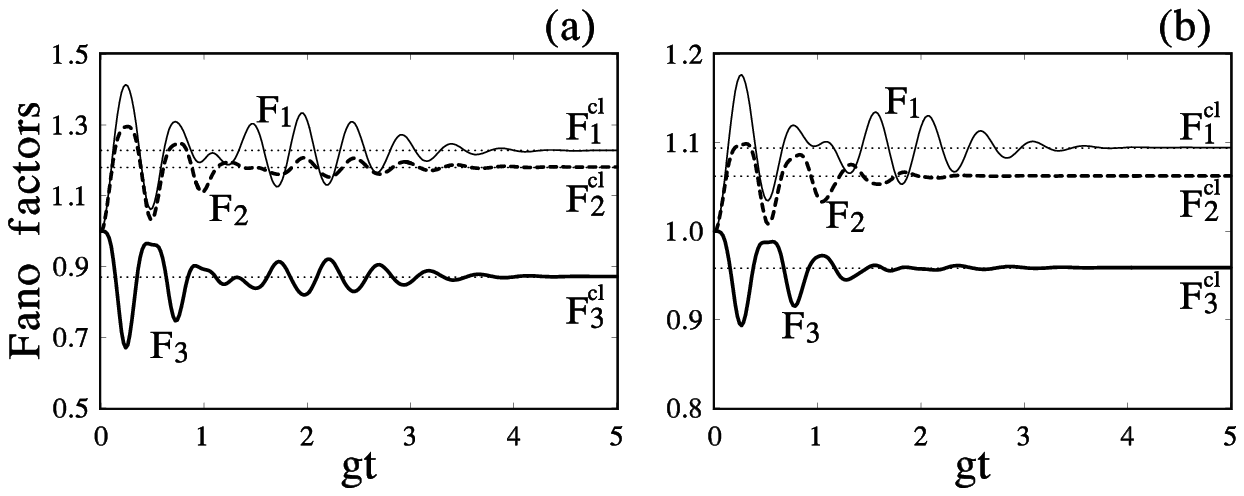}\vspace{0mm}%
\caption{Quantum dynamics {\em in} NETR: Fano factors $F_{1}$
(thin solid), $F_{2}$ (dashed), and $F_{3}$ (thick solid curves)
versus scaled time for different amplitudes of initial coherent
fields: (a) $r_{1}=6$, $r_{2}=4$, and (b) $r_{1}=6$, $r_{2}=2$
together with $r_{3}=r_1r_2/ \sqrt{r_1^2+r_2^2}$. Sub-frequency
modes become super-Poissonian, $F_{1,2}>1$. By contrast, the
sum-frequency mode becomes sub-Poissonian, $F_{3}<1$. Dotted
lines represent the classical trajectory predictions of
$F_{k}^{\rm cl}$ to which all the quantum curves tend
asymptotically.}
\end{figure}
\noindent $Q$-function defined to be
\begin{eqnarray}\label{N04}
Q(\alpha_1,\alpha_2,\alpha_3)&=& \pi^{-3}
\langle\alpha_1,\alpha_2,\alpha_3|\hat{\rho}
|\alpha_1,\alpha_2,\alpha_3\rangle,
\end{eqnarray}
where $|\alpha_1\rangle\otimes|\alpha_2\rangle \otimes|\alpha_3
\rangle$. In figures 1 and 2, we plot its marginal single-mode
Husimi $Q$-functions given by
\begin{eqnarray}\label{N05}
Q(\alpha_k)&=&\int Q(\alpha_1,\alpha_2,\alpha_3)\, \prod_{m\neq
k}{\rm d}^2\alpha_{m},
\end{eqnarray}
where $k,m=1,2,3$. The Fano factors, presented in figures 3 and 4,
were calculated with the help of the marginal $Q$-functions. Due
to obvious computational difficulties, the exact quantum results
can be obtained for relatively small numbers (up to few hundreds)
of interacting photons only.

By analysing the numerical quantum solution we observe that the
basic features of the photon number evolution for the
non-degenerate TWM are in agreement with those for the harmonic
generation processes, as recently reported in
\cite{Bajer99,Bajer00}. In particular, we observe the so-called
no-energy-transfer regime (NETR) \cite{Paul,Bandilla,Drobny}, for
which the energies and intensities of both modes remain constant
in time during the interaction. Although small energy flows
between the modes appear as a consequence of vacuum fluctuations,
their influence is negligible for strong fields. NETR in the
three-wave mixing can be observed if the amplitudes and phases of
the initial coherent fields are matched as follows (Eq. (18) in
\cite{Bandilla}):
\begin{eqnarray}\label{N06}
\frac{1}{r_{3}^{2}}&=&\frac{1}{r_{1}^{2}}+\frac{1}{r_{2}^{2}},
\nonumber \\
\phi _{3} &=& \phi _{1}+\phi _{2}.
\end{eqnarray}
Usually, i.e., for the initial coherent fields {\em not}
satisfying (\ref{N06}), all the Fano factors are stabilized in the
super-Poissonian statistics after a short ($gt\left| r_{k}\right|
\lesssim 1$) relaxation period. Thus, the outputs have high-level
photocount noise. In figure 1, we present a typical quantum
evolution of the single-mode Husimi functions $Q(\alpha_1)$ and
$Q(\alpha_1)$ for the initial amplitudes $r_{1}=6$, $r_{2}=4$ and
$r_{3}=0$. The corresponding evolutions of the photon numbers and
Fano factors are presented in figure 3.

Different behaviour is observed if the initial phase $\phi_3$ and
amplitude $r_{3}$ of the sum-frequency mode fulfill the condition
for NETR. This distinction is clearly seen by comparing figures 1
and 2 for $Q$-functions or figures 3 and 4 for the Fano factors.
In figure 4, the Fano factors are calculated for two different
pairs of the initial amplitudes of sub-frequency modes: (a)
$r_{1}=6,\;r_{2}=4$ and (b) $r_{1}=6,\;r_{2}=2$ and the
sum-frequency-mode amplitude $r_{3}$ fulfilling (\ref{N06}). We
observe that all the Fano factor curves start at $F{}_{k}\left(
0\right) =1$ and after some relaxations become stationary at much
lower noise levels than those for fields
\begin{figure}
\vspace*{-2mm} \hspace*{-7mm} \epsfxsize=9cm\epsfbox{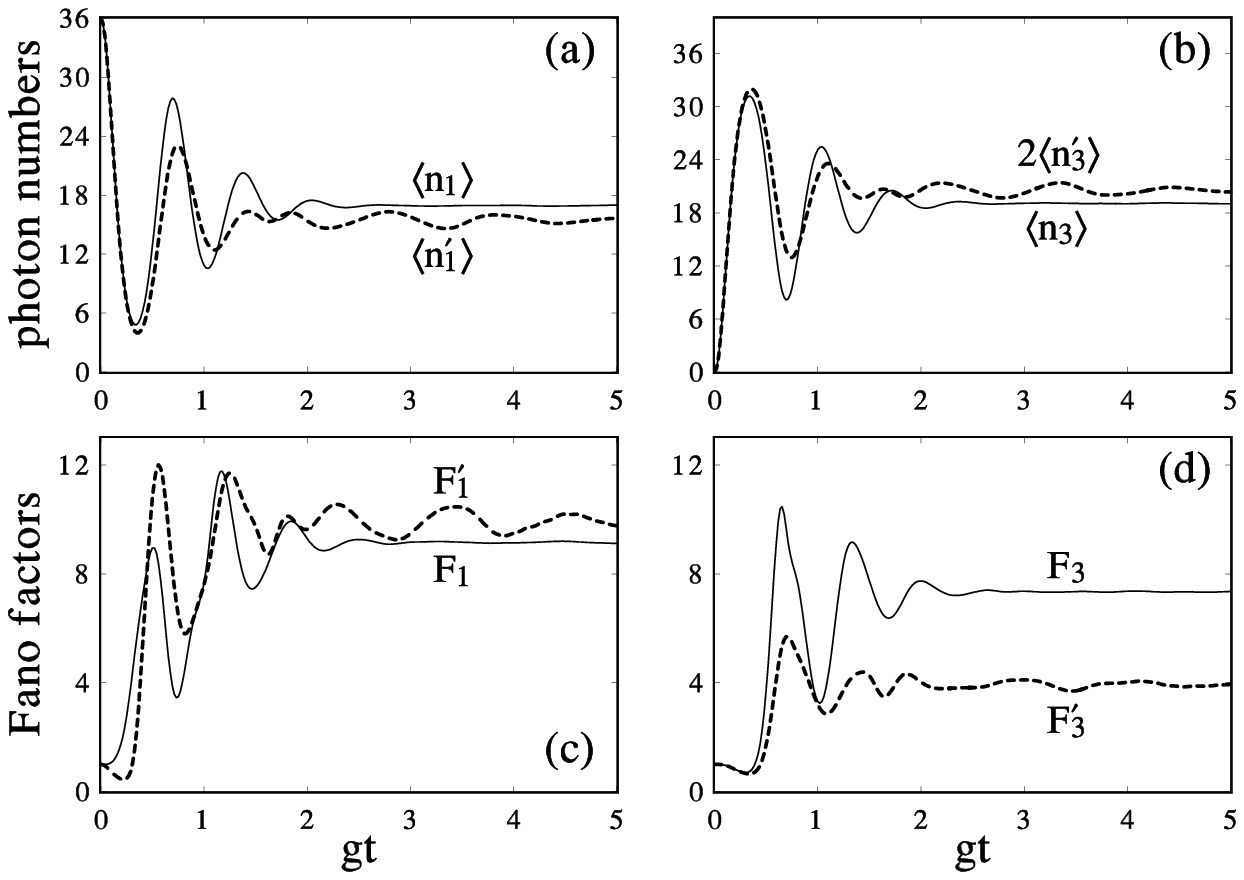}%
\caption{Non-degenerate vs degenerate TWM {\em out} of NETR:
Photon numbers $\langle\hat{n}_k\rangle$ and Fano factors $F_k$
($k=1,3$) are obtained for initial amplitudes $r_1=r_2=6$ and
$r_3=0$, whereas $\langle\hat{n}'_k\rangle$ and $F'_k$ are for
$r'_1=6$ and $r'_3=0$. All quantities with (without) prime
correspond to the degenerate (non-degenerate) TWM.}
\end{figure}
\begin{figure}
\vspace*{-4mm} \hspace*{-7mm} \epsfxsize=9cm\epsfbox{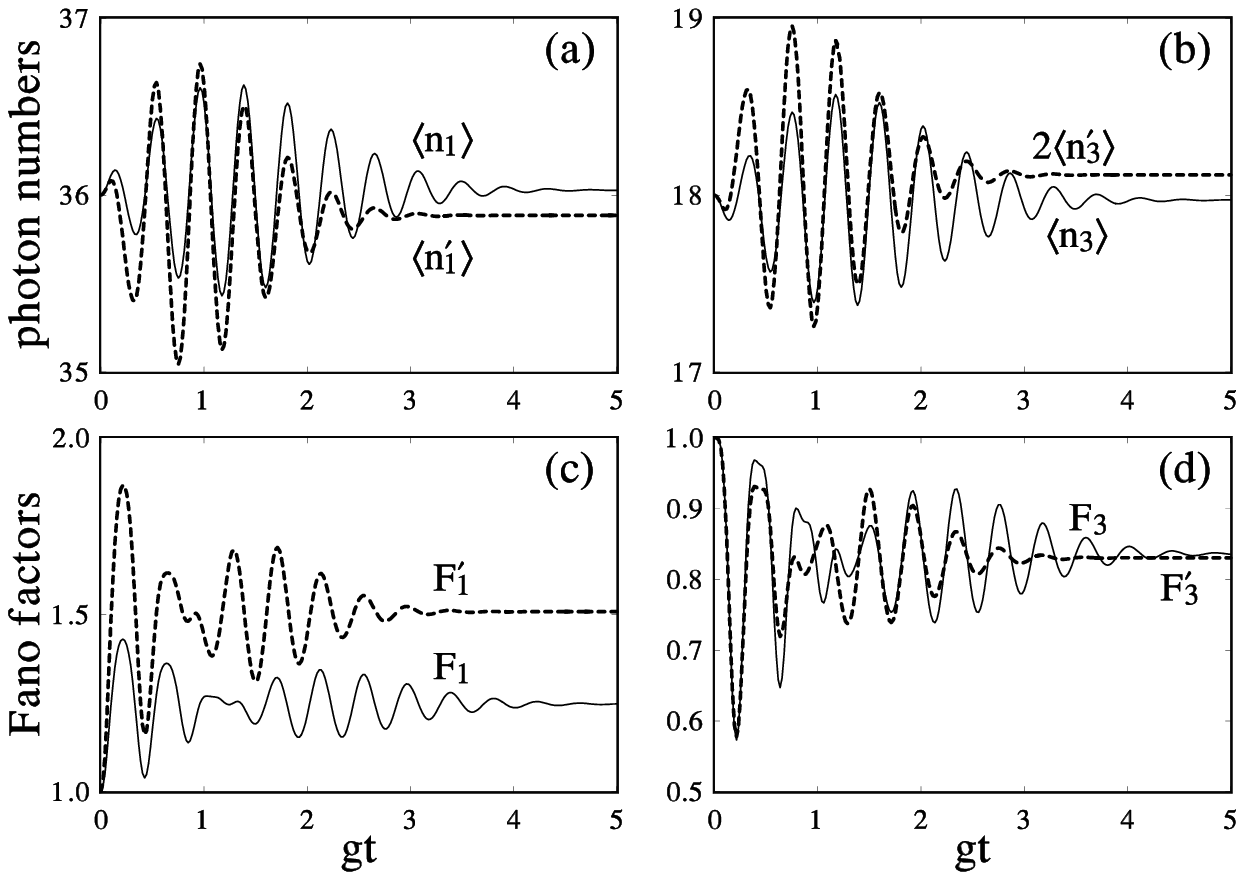}%
\caption{Non-degenerate vs degenerate TWM {\em in} NETR:
$\langle\hat{n}_k\rangle$ and $F_k$  are calculated for initial
amplitudes $r_1=r_2=6$ and $r_3=6/\sqrt{2}$, whereas
$\langle\hat{n}'_k\rangle$ and $F'_k$ are for $r'_1=6$ and
$r'_3=3$.}
\end{figure}
\begin{figure}
\vspace*{-3mm} \hspace*{-7mm} \epsfxsize=9cm\epsfbox{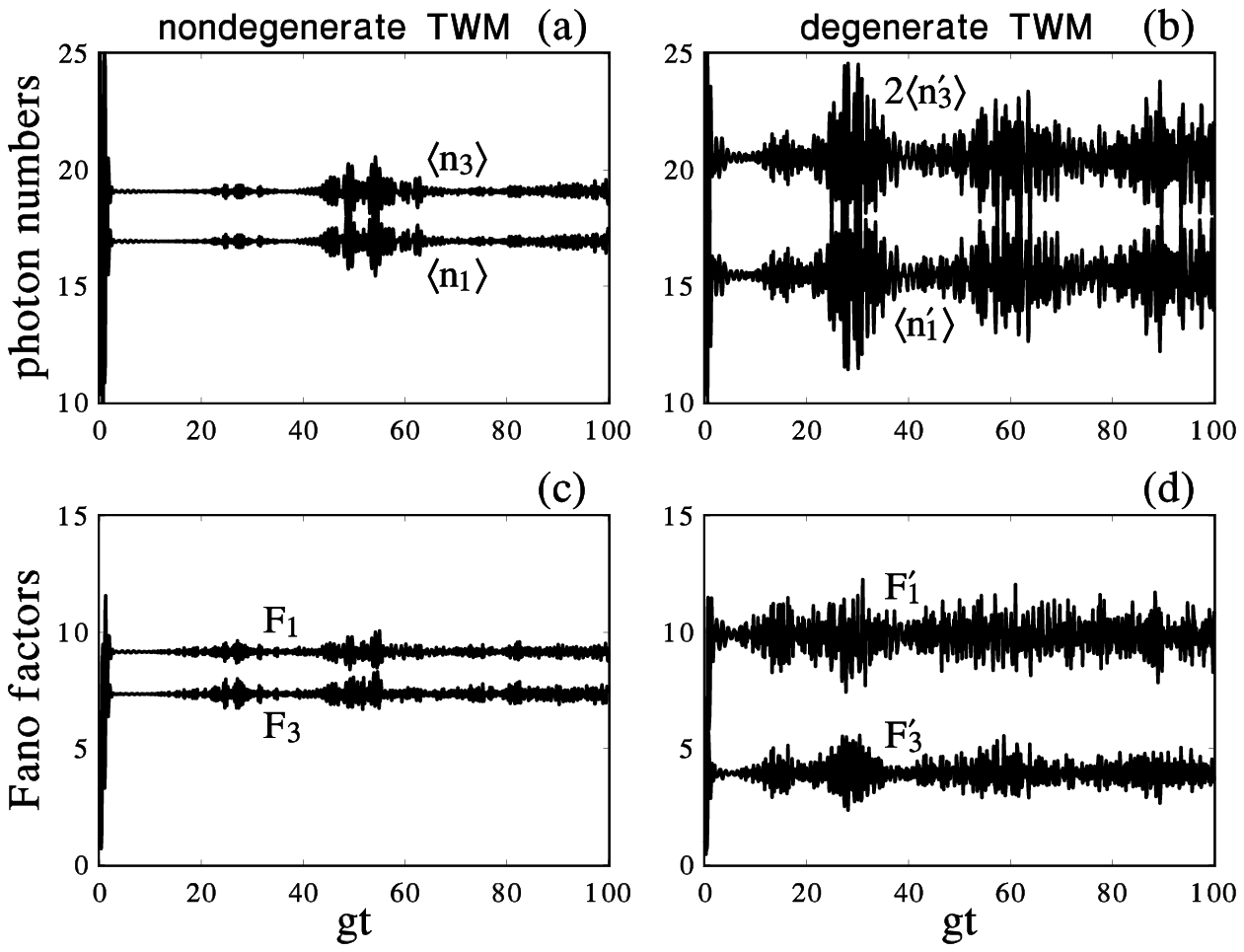}%
\caption{Revivals and collapses in degenerate TWM vs those in
non-degenerate TWM {\em out} of NETR.  Initial conditions are the
same as in figure 5.}
\end{figure}
\begin{figure}
\vspace*{-5mm} \hspace*{-7mm} \epsfxsize=9cm\epsfbox{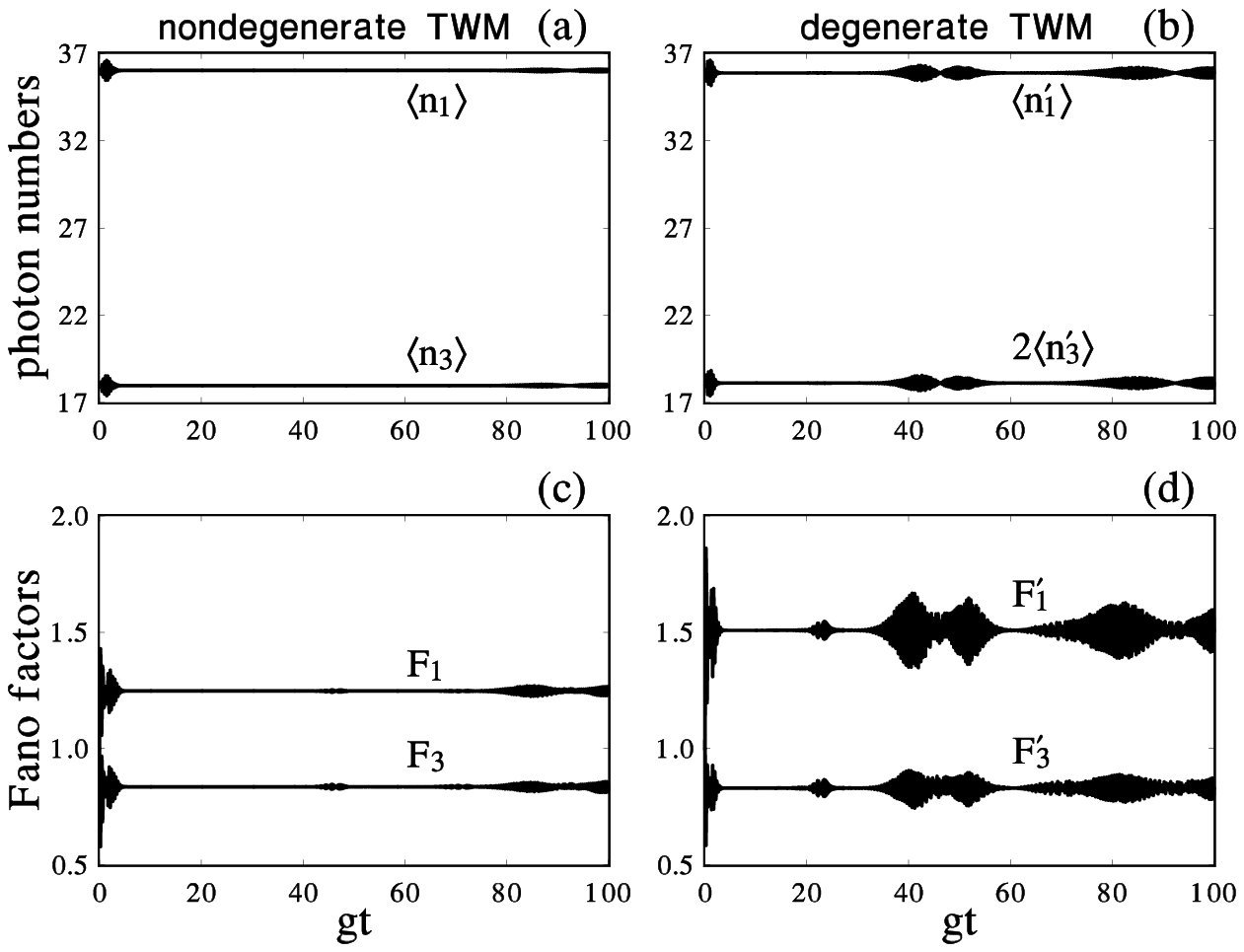}%
\caption{Revivals and collapses in degenerate and non-degenerate
TWM {\em in} NETR with the same initial conditions as those in
figure 6.}
\end{figure}
\noindent%
out of NETR. The sub-frequency modes remain super-Poissonian with
$F{}_{1,2}\left( t\right) >1$, whereas the sum-frequency mode
becomes sub-Poissonian with $F{}_{3}\left( t\right) <1$. The most
suppressed noise is observed for the balanced inputs, given by
$r_{1}=r_{2}$ and $r_{3}=r_{1}/\sqrt{2}$, as a special case of
condition (\ref{N06}). For those inputs, the Fano factor in the
time limit tends to $F{}_{3}\left( t\rightarrow \infty \right)
\approx 0.83$.

As we have shown in Refs.~\cite{Bajer99,Bajer00}, the same degree
of the Fano factor can be obtained in degenerate TWM. Thus, one
can address the following objection: What is the point to study
the same parameter in the closely related non-degenerate version?
First, we stress that the same Fano factor is obtained in a
special case only: for the sum-frequency mode in NETR for
long-interaction times and high-intensity fields.  By contrast,
these factors are distinct for the sub-frequency modes under the
same initial conditions. In general, the results even for the
sum-frequency mode in these two processes are different for
initial conditions either not fulfilling (\ref{N06}) or for lower
intensity fields or different time periods. Second, we will show
that the Fano factors for non-degenerate TWM are better
stabilized than those for degenerate TWM for much longer
evolution times. This is an important advantage of the
non-degenerate conversion.

For better comparison, let us analyse in detail the degenerate TWM
described by
\begin{equation}
\hat{H}'=\hbar g' \left( \hat{a}^2_{1}\hat{a}_{3}^{\dag
}+\hat{a}_{1}^{\dag 2}\hat{a}_{3}\right) ,
\label{N01a}
\end{equation}
For clarity, observables calculated for the degenerate TWM are
marked with prime to distinguish them from those for model
(\ref{N01}) and we keep subscript `3' (not `2') for the
sum-frequency mode.  Hamiltonian (\ref{N01a}) formally differs
from (\ref{N01}) in the assumption of $\hat{a}_{1}=\hat{a}_{2}$
only. But we also put $g'=g/\sqrt{2}$  for better synchronization
of oscillations in photon numbers $\langle \hat{n}'_k\rangle$ and
$\langle \hat{n}_k\rangle$. Sub-Poissonian statistics in
degenerate TWM was analysed in our former works
\cite{Bajer99,Bajer00}. In figures 5 and 6, we compare exact
quantum evolutions of the mean photon numbers and Fano factors
for degenerate ($\langle \hat{n}'_k\rangle$ and $F'_k$ for
$k=1,3$) and non-degenerate ($\langle \hat{n}_k\rangle$ and
$F_k$) TWM for the same initial conditions.  We observe similar
behavior for short times $gt\ll 1$ only.  For longer times (also
for $gt<1$) results are not equal by no means.  Different
predictions of quantum evolutions for models (\ref{N01}) and
(\ref{N01a}) come from different commutation relations:
$[\hat{a}_1,\hat{a}_2^{\dag}]=0$ for non-degenerate TWM and
$[\hat{a}_1,\hat{a}_2^{\dag}]\equiv
[\hat{a}_1,\hat{a}_1^{\dag}]=1$ for degenerate TWM.  As a result,
constants of motion are different:  $\hat{N}'_{\rm
total}(t)\equiv \hat{n}'_1(t) +2\hat{n}'_3(t)$=const for
degenerate TWM, whereas $\hat{n}_1(t)- \hat{n}_2(t)$=const and
$\hat{n}_1(t)+\hat{n}_3(t)$=const for non-degenerate TWM.  We
note that $\hat{N}_{\rm total}(t) \equiv \hat{n}_1(t)+
\hat{n}_2(t)+\hat{n}_3(t) \neq$const in the latter process.
Evolutions, presented in figures 5 and 6 for  degenerate and
non-degenerate TWM, are distinct in amplitudes and frequency of
oscillations as well as in the level of their ``saturation''. For
example, the limiting value of the sum-frequency Fano factor for
degenerate TWM is lower than that for non-degenerate TWM for
evolution out of NETR (see figure 5(d)). While the sub-frequency
Fano factor in NETR is considerably higher for degenerate
compared to non-degenerate TWM (see figure 6(c)).

In figures 1--6, we have analysed the time regime which is long
compared to the typical interaction times for known crystal
lengths.  However, the time is short compared to the revival
times for such systems. The question arises about the
photon-number noise suppression on such a long scale.  This
analysis will show an advantage of non-degenerate over degenerate
TWM related to the stabilization of the suppressed photon-number
noise. In figures 7 and 8, we present the long-time evolution for
$0<gt<100$  of the mean photon numbers and Fano factors for both
non-degenerate ($\langle\hat{n}_k\rangle$ and $F_k$) and
degenerate ($\langle \hat{n}'_k\rangle$ and $F'_k$) TWM.  We
observe that the revivals are strongly pronounced for (i)
degenerate TWM (right figures 7 and 8) compared to non-degenerate
process (left figures), and (ii) outside NETR (figure 7) rather
than in NETR (figure 8).  Thus, the non-degenerate TWM in NETR
exhibits the highest stability. Even for longer evolution times
as $100<gt<1000$ of non-degenerate TWM, the oscillations are
similar to those for $0<gt<100$ and it is hardly difficult to
classify them as a typical revival. Oscillations in
$\langle\hat{n}_k\rangle$ are of order $10^{-3}$ and in $F_k$ are
of order $10^{-2}$ even for such small initial amplitudes equal to
$\alpha_1=\alpha_2=6$ and $\alpha_3=6/\sqrt{2}$.  Our analysis is
restricted to initial coherent inputs. It is worth noting that
the revivals are much stronger for quantum input fields like,
e.g., Fock states. In conclusion, although the degenerate and
non-degenerate TWM lead to approximately the same photon-number
noise suppression in the sum-frequency mode for NETR (see figure
6(d)), the non-degenerate process offers much better stabilization
of the suppressed noise for long evolution times (compare figures
8(c) and 8(d)).

In the next sections, we will apply an approximate method of
classical trajectories to explain the extraordinary stabilization
of the observed photocount noise and to estimate analytically the
level of noise suppression for NETR.

\section{Classical analysis}

Complete quantum solution of the model given by Hamiltonian
(\ref{N01}) can be found numerically only. Yet, in a special case
for strong fields, analytical results can be obtained by applying
approximate classical methods.

In analogy to quantum Hamiltonian (\ref{N01}), the classical model
of non-degenerate TWM can be described by \cite{Boyd}:
\begin{equation} \label{N07}
{\cal H}= g\left( \alpha_{1}\alpha_{2}\alpha_{3}^{*}+ {\rm
c.c.}\right),
\end{equation}
where $\alpha_{k}$ are the complex amplitudes of the $k$-th mode
and $g$ is a non-linear coupling parameter. From (\ref{N07}), one
readily obtains the following complex differential equations
\begin{eqnarray} \label{N08}
\dot{\alpha _{1}} &=&-{\rm i}g\alpha _{2}^{\ast }\alpha _{3},
\nonumber \\
\dot{\alpha _{2}} &=&-{\rm i}g\alpha _{1}^{\ast }\alpha _{3},
\nonumber \\
\dot{\alpha _{3}} &=&-{\rm i}g\alpha _{1}\alpha _{2}.
\end{eqnarray}
It is easy to show by comparing (\ref{N08}) with equations (12)
and (13) from Ref.  \cite{Bajer99} that the classical models for
the degenerate and non-degenerate TWM are equivalent for
$\alpha_1=\alpha_2$ and arbitrary evolution times. To get
equations of motion for degenerate TWM, it is enough to replace
$g$ by $\sqrt{2} g'$ and $\alpha_3$ by $\sqrt{2}\alpha'_3$ in
(\ref{N08}). As was discussed in the former section, the quantum
evolutions of degenerate and non-degenerate TWM are equivalent
for $gt\ll 1$ only.

By introducing real amplitudes and phases, $\alpha
_{k}=r_{k}e^{{\rm i}\phi _{k}}$, equations (\ref{N08}) can be
transformed into the following four real equations
\begin{eqnarray} \label{N09}
\dot{r}_{1} &=&-gr_{2}r_{3}\sin \theta ,  \nonumber \\
\dot{r}_{2} &=&-gr_{1}r_{3}\sin \theta ,  \nonumber \\
\dot{r}_{3} &=&gr_{1}r_{2}\sin \theta ,  \nonumber \\
\dot{\theta} &=& g\left(
\frac{r_{1}r_{2}}{r_{3}}-\frac{r_{1}r_{3}}{r_{2}}-
\frac{r_{2}r_{3}}{r_{1}}\right) \cos \theta ,
\end{eqnarray}
where $\theta =\phi _{1}+\phi _{2}-\phi _{3}$ is the phase
mismatch. The system (\ref{N09}) has three integrals of motion
\begin{eqnarray} \label{N10}
E_{1} &=&r_{1}^{2}+r_{3}^{2}=n_{1}+n_{3},
\nonumber \\
E_{2} &=&r_{2}^{2}+r_{3}^{2}=n_{2}+n_{3},
\nonumber \\
K &=&r_{1}r_{2}r_{3}\cos \theta .
\end{eqnarray}
By extracting $r_{1},r_{2}$ and $\theta $ from (\ref{N09}),
equation for the remaining amplitude $r_{3}$ reads as
\begin{eqnarray}\label{N11}
\left( r_{3}\dot{r}_{3}/g\right) ^{2}+K^{2}=r_{3}^{2}\left(
E_{1}-r_{3}^{2}\right) \left( E_{2}-r_{3}^{2}\right)
\end{eqnarray}
or, equivalently, for the intensity $n_{3}=r_{3}^{2}$ as
\begin{eqnarray} \label{N12}
\left( \dot{n}_{3}/2g\right) ^{2}&=&n_{3}\left( E_{1}-n_{3}\right)
\left( E_{2}-n_{3}\right) -K^{2} \nonumber \\
&=&\left(a-n_{3}\right) \left( b-n_{3}\right) \left(
n_{3}-c\right),
\end{eqnarray}
where the numbers $a\geq b\geq c$ are the roots of cubic equation
$n_{3}\left( E_{1}-n_{3}\right) \left( E_{2}-n_{3}\right)
-K^{2}=0$ satisfying the conditions $abc=K^{2}, \;a+b+c=E_{1}
+E_{2}$, and $ab+ac+bc=E_{1}E_{2}$. Then, the solution for
$n_{3}\left( t\right) $ is found to be
\begin{eqnarray}\label{N13}
n_{3}\left( t\right) =c+\left( b-c\right) {\rm {sn}}^{2}\left[
\sqrt{a-c} gt+\phi _{0},k\right] ,
\end{eqnarray}
where ${\rm {sn}}(u,k)$ is the Jacobi elliptic function with
$k=\sqrt{\frac{b-c}{a-c}}$ and $\phi _{0}$ is the initial phase
given by the elliptic integral of the first kind
\begin{eqnarray}\label{N14}
\phi _{0}=F\left( z,k\right) =\int_{0}^{z} \frac{dx}
{\sqrt{1-k^{2} \sin ^{2}x}},
\end{eqnarray}
where $z=\arcsin\sqrt{(n_{3}(0)-c)/(b-c)}$. One observes that
$n_{3}$ is a periodic function oscillating between the values $c$
and $b$ with the period given by $T=4F\left( \frac{\pi
}{2},k\right)/g$.

In two special cases, solution (\ref{N13}) reduces to the
elementary solutions:
\begin{eqnarray}\label{N15}
n_{3}\left( t\right) =r^{2}{\rm tanh}^{2}(r gt)
\end{eqnarray}
for $r_{1}=r_{2}=r$ and $r_{3}=0$, and
\begin{eqnarray}\label{N16}
n_{3}\left( t\right) =r^{2}{\rm sech}^{2}(r gt)
\end{eqnarray}
for $r_{1}=r_{2}=0$ and $r_{3}=r$.  Another elementary solution is
obtained for the initial fields fulfilling conditions (\ref{N06}).
In this case, the solution reads as ($k=1, 2, 3$)
\begin{eqnarray} \label{N17}
\alpha _{k}\left( t\right)  &=& r_{k}\exp \left(-{\rm i} \frac{r_1
r_{2}r_{3}}{r_{k}^{2}}\, gt\right),
\end{eqnarray}
which describes the classical {\em no-energy-transfer regime}
(NETR) \cite{Paul}, since the amplitudes and energies of all
interacting modes remain constant, i.e., $n_{k}(t) =|\alpha
_{k}(t)| ^{2}=r_{k}^{2}$.  We conclude that NETR observed in the
quantum numerical analysis presented in former section
corresponds to the classical solution (\ref{N17}).

\section{Classical trajectory analysis}

Classical solutions, as presented in the preceding section, do not
describe quantum noise. Nevertheless, they can be used for
simulation of quantum noise if the initial complex amplitudes are
chosen randomly. This approach, referred to as the method of {\em
classical trajectories}, has been applied successfully in a
description of noise in various quantum-optical phenomena
\cite{Bajer99,Bajer00,Bandilla1,Chmela,Nikitin,Bandilla,Drobny,%
Paul,Milburn,Sand}. By analysing $Q$-functions and Fano factors,
we will show that the method of classical trajectories properly
simulates photocount noise in the TWM processes.

To calculate statistical moments, like the Fano factors, one needs
to analyse  the classical evolution of each process (trajectory)
separately and then to average the moments over all the obtained
trajectories. The classical Fano factor, defined to be
\begin{eqnarray}\label{N18}
F^{\rm cl}=\frac{\overline{n^{2}}-\bar{n}^{2}}{\bar{n}},
\end{eqnarray}
can be obtained by the classical trajectory averaging. We denote
this averaging by bar to distinguish it from quantum ensemble
averaging indicated by brackets. We will apply the method of
classical trajectories along the lines of the analysis presented
in \cite{Bajer99}. We choose the initial amplitudes to be $\alpha
_{k}=r_{k}$ and blur them with the Gaussian noise, which results
in
\begin{equation} \label{N19}
\alpha _{k}=r_{k}+x_{k}+{\rm i}y_{k},
\end{equation}
where $x_{k}$ and $y_{k}$ are mutually independent real Gaussian
stochastic quantities with the identical variances $\sigma
^{2}=1/4$. We assume further that the fields are strong, i.e.,
$r_{k}\gg 1$.

By substituting (\ref{N19}) into (\ref{N10}), the integrals of
motion can be expressed as
\begin{eqnarray} \label{N20}
E_{1} &=&\left| \alpha _{1}\right| ^{2}+\left| \alpha _{3}\right|
^{2}=r_{1}^{2}+r_{3}^{2}+c_{1},
\nonumber \\ E_{2} &=&\left|
\alpha _{2}\right| ^{2}+\left| \alpha _{3}\right|
^{2}=r_{2}^{2}+r_{3}^{2}+c_{2},
\nonumber \\
K &=& {\rm Re} \left( \alpha _{1}\alpha _{2}\alpha _{3}^{\ast
}\right) = r_{1}r_{2}r_{3}+d_{1},
\end{eqnarray}
where $c_{1},$ $c_{2}$ and $d_{1}$ are small corrections of
$E_{1}$, $E_{2}$ and $K$, respectively, depending on $r_{k}$,
$x_{k}$ and $y_{k}$. To eliminate the linear term in the RHS of
(\ref{N12}), we substitute
\begin{eqnarray}\label{N21}
n_{3}&=&\frac{1}{3}E_{2}+\frac{1}{3}E_{1}-\frac{1}{3}\sqrt{\left(
E_{2}^{2}-E_{1}E_{2}+E_{1}^{2}\right) }+\epsilon
\nonumber \\
&=& n_{30}+b+\epsilon ,
\end{eqnarray}
where $\epsilon $ is a small correction of stationary value
$n_{30}+b$. Under the strong-field approximation ($r_{k}\gg 1$),
one can neglect the small correction $\epsilon^{3}$ and the RHS
of (\ref{N12}) can be approximated by the quadratic function
$\Omega ^{2}\left( a^{2}-\epsilon ^{2}\right)$. Thus, one gets
simple equation
\begin{equation} \label{N22}
\dot{\epsilon}^{2}=\left( 2g\Omega \right) ^{2}\left(
a^{2}-\epsilon ^{2}\right),
\end{equation}
which leads to the solution of (\ref{N12}) in the following form
\begin{figure}
\vspace*{-3mm} \hspace*{-3mm}
\epsfxsize=8.5cm\epsfbox{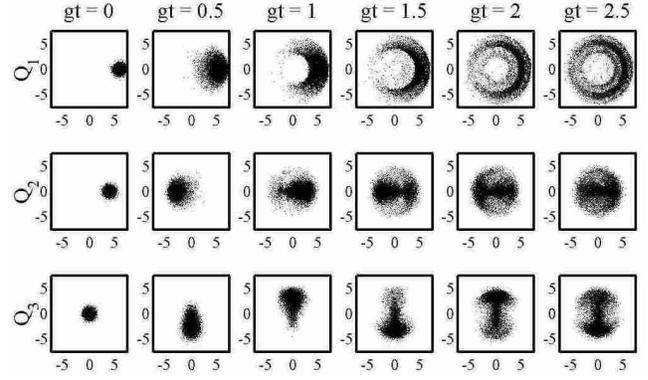}\vspace*{4mm}%
\caption{Classical simulation of typical quantum dynamics {\em
out} of NETR: Clouds of 10,000 points representing marginal
$Q$-functions for the same initial conditions and times as in
figure 1. }
\end{figure}
\begin{figure}
\vspace*{-7mm} \hspace*{-3mm} \epsfxsize=8.5cm
\epsfbox{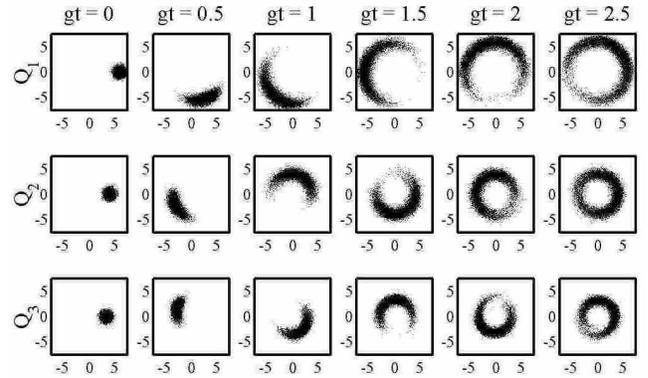}\vspace*{3mm}%
\caption{Classical simulation of quantum dynamics {\em in} NETR
for the same cases as in figure 2.}
\end{figure}
\begin{equation} \label{N23}
n_{3}\left( t\right) =n_{30}+b+a\sin \left( 2g\Omega t+\varphi \right) ,
\end{equation}
where $n_{30}=r_{3}^{2}=r_{1}^{2}r_{2}^{2}/(r_{1}^{2}+r_{2}^{2})$.
The coefficients $a$, $b$ and $\Omega $, together with $c_{1}$ and
$c_{2}$,  are complex functions of $r_{k}$ and noise parameters
$x_{k}$ and $y_{k}$. With the help of integrals of motions, given
by (\ref{N10}), solutions for other modes ($k=1,2$) can readily be
found as
\begin{equation} \label{N24}
n_{k}(t) = E_{k}-n_{3}(t) =r_{k}^{2} +c_{k}-b-a\sin( 2g\Omega
t+\varphi).
\end{equation}
We observe that all the three solutions, given by (\ref{N23}) and
(\ref{N24}), are of the form of large constants slightly perturbed
by the same harmonic function. In figures 9 and 10, we present
classical simulation of quantum dynamics by calculating time
evolutions of 10,000 points in phase space according to classical
equations of motion. These representations correspond to the
Husimi $Q$-functions presented in figures 1 and 2, respectively.
By comparing figures 1 and 2 or, equivalently, figures 9 and 10,
we observe two distinct types of evolution determined by the
initial amplitudes to be in NETR or out of it.

The classical and quantum descriptions are principally different
in detail.  Thus, our plots of the $Q$-function based on the
exact quantum solution of TWM (figures 1 and 2) and those
obtained by an approximate classical simulation (figures 9 and
10) also differ in detail. The discrepancies are more pronounced
for lower amplitude inputs and longer interactions.  Moreover,
the methods of graphical representations are different:  a
topographical picture of $Q_k(\alpha)$ versus a cloud of
classical points.  Nevertheless,  ``it is surprising how close
the clouds of dots are to the $Q$-function''~\cite{Nikitin}. The
clear correspondence between figures 2 and 10 or 1 and 9
justifies our application of the classical trajectory
approximation.

As the next step of the classical trajectory method, one has to
perform averaging of solutions (\ref{N23}) and (\ref{N24}) to
calculate the desired moments. We find that the mean values of the
parameters occurring in solution (\ref{N24}) are $\bar{b}
=\overline{c_{1}}= \overline{c_{2}} =0$ and
\begin{eqnarray} \label{N25}
\bar{\Omega} &=& \sqrt{r_1^2+r_2^2-r_3^2}=
\sqrt{\frac{r_{2}^{4}+r_{1}^{2}r_{2}^{2}+r_{1}^{4}}{
r_{1}^{2}+r_{2}^{2}}}
\end{eqnarray}
together with their mean quadratic moments
\begin{eqnarray} \label{N26}
\overline{b^{2}} &=& \frac{2A}{r_{2}^{2}+r_{1}^{2}}
(2r_{1}^{8}+r_{1}^{6}r_{2}^{2}+2r_{1}^{4}r_{2}^{4}
+r_{1}^{2}r_{2}^{6}+2r_{2}^{8}),
\nonumber\\
\overline{(c_{1}-b)^{2}} &=& \frac{2A}{r_2^2}
(4r_{1}^{8}+10r_{1}^{6}r_{2}^{2}+11r_{1}^{4}r_{2}^{4}
+7r_{1}^{2}r_{2}^{6}+2r_{2}^{8}),
\nonumber\\
\overline{\left( c_{2}-b\right) ^{2}} &=& \frac{2A}{r_2^2}
(2r_{1}^{8}+7r_{1}^{6}r_{2}^{2}+11r_{1}^{4}r_{2}^{4}
+10r_{1}^{2}r_{2}^{6}+4r_{2}^{8}),
\nonumber\\
\overline{a^{2}} &=&2A ( 4r_{1}^{6}+7r_{1}^{4}
r_{2}^{2}+7r_{1}^{2}r_{2}^{4}+4r_{2}^{6}),
\end{eqnarray}
given in terms of the auxiliary function
\begin{eqnarray}\label{N27}
A=\frac{r_1^2r_2^2}{8(r_1^4+r_1^2r_2^2+r_2^4)^2}\,.
\end{eqnarray}
The phase $\varphi$ can be obtained from (\ref{N23}) at $t=0$.
Thus, the photon-number mean values are simply equal to
${\bar{n}_{k}} = r_{k}^{2}$ and their variances are given by
($k=1,2,3$)
\begin{equation} \label{N28}
\overline{n_{k}^{2}}-\bar{n} _{k}^{2} =\overline{\left(
c_{k}-b\right) ^{2}}+\frac{1}{2}\overline{a^{2}}
\end{equation}
in terms of the moments (\ref{N26}) and $c_{3}\equiv 0$. The term
$\overline{\sin ^{2}\left( 2g\Omega t+\varphi \right) }$ has
simply been estimated as $\frac{1}{2}$ for sufficiently large $t$,
when  $\bar{n}_{k}$ and $F_{k}$ become time-independent. Thus, we
arrive at the following Fano factors
\begin{eqnarray} \label{N29}
F_{1}^{\rm cl} &=&1+ A(8r_{1}^{4}+5r_{1}^{2}r_{2}^{2}+5r_{2}^{4}),
\nonumber \\
F_{2}^{\rm cl} &=&1+ A(5r_{1}^{4}+5r_{1}^{2}r_{2}^{2}+8r_{2}^{4}),
\nonumber\\
F_{3}^{\rm cl} &=& 1-3A(r_{1}^{2}+r_{2}^{2})^2,
\end{eqnarray}
where $A$ is given by (\ref{N27}). As one could expect, the
formulas for $F_{k}^{\rm cl}$ are symmetric with respect to
exchange of the subscripts $1\longleftrightarrow 2$. We finally
conclude that the TWM in the no-energy-transfer regime can be a
source of time-stable sub-Poissonian light in the sum-frequency
mode as described by the Fano factor
\begin{figure}
\vspace*{-1cm} \hspace*{0mm} \epsfxsize=7.5cm \epsfbox{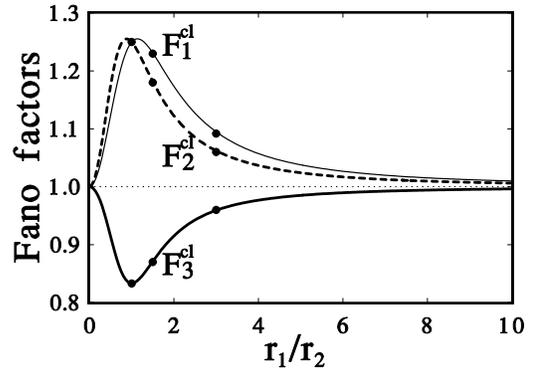}
\vspace*{0mm}%
\caption{Classical predictions of Fano factors $F_{k}^{\rm cl}$
($k=1,2,3$) versus ratio $r_{1}/r_{2}$ of the input coherent-field
amplitudes. Small circles represent the quantum Fano factors
$F_{k}$ obtained from the exact quantum solutions presented in
figures 4 and 6.} \end{figure}%
\begin{eqnarray} \label{N30}
F_{3}^{\rm cl}(\rho) &=& 1-\frac{3\rho(1 + \rho)^2}{8(1 + \rho +
\rho^2)^2}\le 1,
\end{eqnarray}
depending on ratio of the mean intensities of initial coherent
fields defined by $\rho=r^2_{1}/ r^2_{2}$  if $r_1>0$ or
$\rho=r^2_{2}/r^2_{1}$ if $r_2>0$. The sub-frequency fields become
super-Poissonian with the Fano factors
\begin{eqnarray} \label{N31}
F_{1}^{\rm cl}(\rho) =F_{2}^{\rm cl}(1/\rho) = 1+ \frac{\rho(5 +
5\rho + 8\rho^2)}{8(1+\rho+\rho^2)^2}\ge 1.
\end{eqnarray}
In figure 11, the classical predictions of the Fano factors are
depicted as a function $r_1/r_2$. By analysing (\ref{N30}) and
figures 4, 6, and 11, we conclude that the sum-frequency mode
solely is sub-Poissonian ($F_{3}^{\rm cl}\leq 1$) and the
strongest noise suppression is obtained for
$r_{1}=r_{2}=r_{3}/\sqrt{2}$, when $F_{3}^{\rm cl}=5/6\approx
\allowbreak 0.\,\allowbreak 833$. For highly unbalanced input
intensities $r_{1}\ll r_{2}$ or $r_{1}\gg r_{2}$, all the Fano
factors approach unity $F_{1}^{\rm cl}\approx F_{2}^{\rm
cl}\approx F_{3}^{\rm cl}\rightarrow 1$. Mutually equal Fano
factors, estimated by $F_{1}^{\rm cl}=F_{2}^{\rm
cl}=5/4=\allowbreak 1.\,\allowbreak 25$, are predicted for the
balanced inputs $r_{1}=r_{2}=r_{3}/\sqrt{2}$. The maximum values
of the Fano factors, estimated by $F_{1}^{\rm cl}=F_{2}^{\rm
cl}={\rm max}=1.\,\allowbreak 255$, are obtained for slightly
unbalanced inputs: $r_{1}=1.\,\allowbreak 136 r_{2}$  to maximize
$F_{1}^{\rm cl}$ and for $r_{1}=0.\,\allowbreak 881 r_{2}$ to
maximize $F_{2}^{\rm cl}$.

We have predicted in \cite{Bajer99,Bajer00} the stationary
sub-Poissonian Fano factors for the second ($F^{\rm cl}_{2}=5/6$)
and third ($F^{\rm cl}_{3}=13/16$) harmonic generations within
NETR.  The minimum value of the sum-frequency-mode Fano factor
for the non-degenerate TWM is the same as that obtained for
degenerate TWM (i.e., second-harmonic generation) \cite{Bajer99},
but higher than that for degenerate four-wave mixing (i.e.,
third-harmonic generation). However, for the sub-frequency modes,
the Fano factors for the non-degenerate TWM are smaller than
those for the degenerate cases, namely $F_{1,2}^{\rm cl}(1)=5/4$
instead of $3/2$ and $29/16$, respectively.

In figures 3 and 4, we have compared evolutions of the exact
quantum Fano factors $F_k$ (depicted by solid or dashed curves)
with their classical estimations, $F^{\rm cl}_k$ (dotted lines).
One observes that all the curves start at $F_{k}=1$ for initial
coherent fields and after some relaxations become quasi-stationary
with $F_k\approx F_{k}^{\rm cl}$, given by (\ref{N29}). It is
worth noting that very good estimation is achieved even for
relatively small amplitudes, e.g., $r_{k}\leq 6$. We conclude that
the conditions for NETR in quantum dynamics and suppression of
the observed quantum noise levels are well explained by the
classical trajectory method.

Finally, we will compare scaling properties of the Fano factors in
their dependence on light intensity and initial amplitude for the
degenerate and non-degenerate TWM. Drobn\'y et al. \cite{Drobny}
calculated the scaling laws under the truncated Wigner
approximation for the maximum sub-Poissonian photon-number noise
in TWM.  Their formulas are valid also in the limit of
$r=\alpha_1(0)\rightarrow \infty$. Here, we focus on nonlinear
fits for finite ranges of $r$ only. Let us investigate the
scaling properties of the maximum sub-Poissonian character of the
sum-frequency mode corresponding to the first minimum of the
$F_3$ and $F_3'$ curves in figures 3(b), 4, 5(d), and 6(d).  In
figure 12, we plot the exact quantum numerical values of $\min_t
F_3(x,t)\equiv F_3(x,t_{\min})$ or $\min_t F'_3(x,t)\equiv
F'_3(x,t'_{\min})$ as a function of initial amplitudes
$x=r_1=r_2=r'_1$ and of intensities $x\equiv
\langle\hat{n}_3(t_{\min})\rangle =\langle\hat{n}'_3
(t'_{\min})\rangle$. We fit those minima with the exponent and
polynomial functions of parameters listed in tables I and II. The
errors, given in the last column, are estimated by the standard
deviation.  We observe that the scaling laws $a x^b$ for non-degenerate%
\vspace*{-1mm}
\begin{table}
\caption{Scaling laws $a x^b$ for non-degenerate TWM ($\min_t
F_3(x)$) and degenerate TWM ($\min_t F'_3(x)$) {\em out} of NETR.
Initial conditions are $1<r_1=r_2\le 10$, $1<r'_1\le 20$, and
$r_3=r_3'=0$.}
\begin{center}
\begin{tabular}{clccc}
No. & fitted function & $a$    & $b$     & error \\
\hline
1 & $\min_t F_3(r_1)$ & 0.8819  & -0.1254  & 0.0004\\
2 & $\min_t F_3(\langle\hat{n}_3\rangle)$  & 0.8560  & -0.0572  & 0.0001\\
3 & $\min_t F'_3(r_1')$ & 0.7694  & -0.0906  & 0.0003\\
4 & $\min_t F'_3(\langle\hat{n}'_3\rangle)$  & 0.7352  & -0.0427  & 0.0003\\
\end{tabular}
\end{center}
\end{table}
\vspace*{-9mm}
\begin{table}
\caption{Polynomial fit $(a x^2+b x+c)/x^2$ for non-degenerate
($\min_t F_3(x)$) and degenerate ($\min_t F'_3(x)$) TWM {\em in}
NETR. Initial conditions are $1<r_1=r_2\le 10$, $1<r'_1\le 20$,
$r_3=r_1/\sqrt{2}$, and $r_3'=r'_1/2$. }
\begin{center}
\begin{tabular}{clccccc}
No. & fitted function & $r_1$ & $a$ & $b$ & $c$  & error\\
\hline
1 & $\min_t F_3(r_1)$ & $>1$ & 0.5474 & 0.1104 & 0.2956 & 0.0003\\
 &      & $\gg 1$ & 0.5519 & 0.0643 & 0.3894 & 0.0007\\
2 & $\min_t F_3(\langle\hat{n}_3\rangle)$
      & $>1$ & 0.5562 & 0.3143 & -0.1618 & 0.0001\\
 &      & $\gg 1$ & 0.5559 & 0.3240 & -0.2057 & 0.0008\\
3 & $\min_t F'_3(r_1')$ & $>1$ & 0.5495 & 0.1195 & 0.3510 & 0.0003\\
 &      & $\gg 1$ & 0.5541 & 0.0406 & 0.5739 & 0.0012\\
4 & $\min_t F'_3(\langle\hat{n}'_3\rangle)$
      & $>1$ & 0.5558 & 0.2057 & -0.0607 & 0.0001\\
 &      & $\gg 1$ & 0.5556 & 0.2121 & -0.0783 & 0.0005\\
\end{tabular}
\end{center}
\end{table}
\begin{figure}
\vspace*{-3mm} \hspace*{-7mm} \epsfxsize=9cm\epsfbox{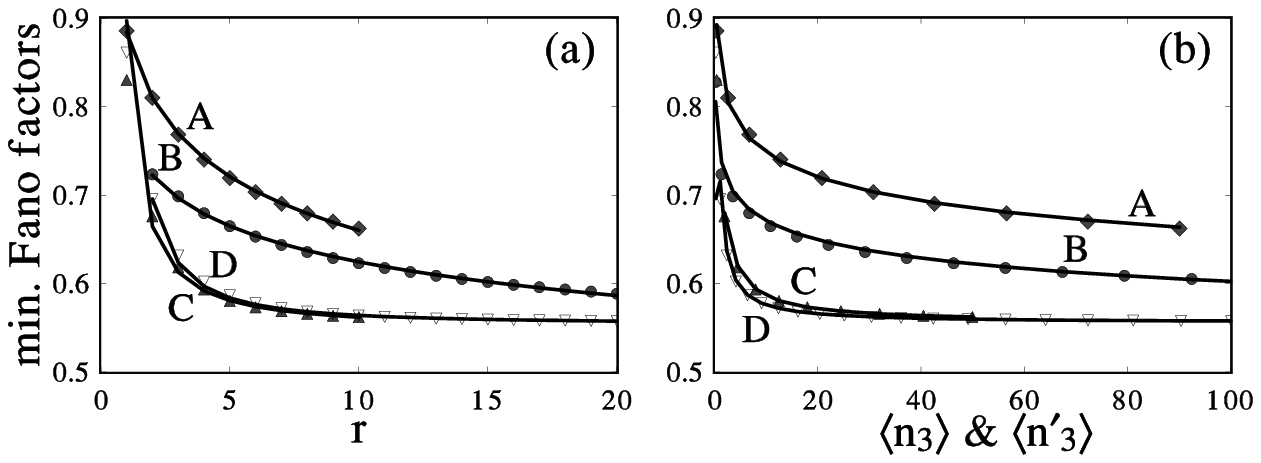}%
\caption{Maximum photon-number noise suppression for
non-degenerate (curves A and C) and degenerate  (B and D) TWM:
Time-minimized Fano factors, $\min_t F_3$ and $\min_t F'_3$, as a
function of (a) initial amplitudes $r\equiv r_1 =r'_1$ and (b)
mean photon numbers $\langle\hat{n}_3\rangle
=\langle\hat{n}'_3\rangle$. Curves: A (with diamonds) for initial
amplitude $r_3=0$, B (with circles) for $r'_3=0$, C (with solid
triangles) for $r_{3}=r_1/\sqrt{2}=r/\sqrt{2}$, D (with empty
triangles) for $r'_3=r'_1/2=r/2$. The marked points are obtained
from the exact quantum solutions and are fitted with the
functions given in tables I and II. }
\end{figure}
\noindent%
and degenerate TWM give good approximation of the exact values
{\em out} of NETR only. However, the $a x^b$ law fails to
describe with good precision $\min_t F_3(x)$ and $\min_t F'_3(x)$
{\em in} NETR at least for intensities up to 100 photons.  Thus
instead of the exponent law we apply the (inverse) polynomial fit
of the form $(a x^2+b x+c)/x^2$, where we introduce $x^{-2}$ in
relation to the definition of Fano factor. By contrast to the
exponent fits, the polynomial laws give very good predictions of
the maximum sub-Poissonian behavior at least for $1<r_1=r_2\le
10$ and $1<r'_1\le 20$ as seen in figure 12 and table II. The
scaling laws of Drobn\'y et al. \cite{Drobny} differ slightly
from ours presented in table I. The minor differences in the
fitted parameters result from different ranges of $r$ used in the
fitting procedures and from application of the truncated Wigner
approximation in Ref. \cite{Drobny} compared to our exact quantum
method.

On the other hand, the Fano factors for the balanced ($r_1=r_2$)
non-degenerate TWM under NETR conditions for long times and high
intensity fields do not depend on light intensity, which follows
from equations (\ref{N30}) and (\ref{N31}). Similarly, there are
no scaling properties of the Fano factors in the degenerate TWM
for long times and high intensity fields in NETR, i.e., under the
same conditions as those assumed in our classical trajectory
analysis.

\section{Conclusions}

We have analysed the long-time interactions in the non-degenerate
three-wave mixing. To the best of our knowledge, our quantum
analysis is the first presentation of the exact and completely
quantum solution of the nondegenerate three-wave mixing.  In
literature, a special solution can be found for initial
sub-frequency fields with zero amplitudes $\alpha_1=\alpha_2=0$
only. The no-energy-transfer regime for proper choices of
amplitudes and phases of the initial coherent fields has been
observed. We have compared the evolutions of the Husimi
$Q$-functions and their classical trajectory simulations for
processes in the no-energy-transfer regime and out of it. We have
shown numerically in the quantum-mechanical approach that the
three-wave mixing in the no-energy-transfer regime exhibits the
time-stable photocount statistics. This phenomenon was explained
analytically by applying the method of classical trajectories. We
have shown that the sub-frequency modes become super-Poissonian
with the Fano factor $F_{1,2}>1$, whereas the sum-frequency mode
becomes sub-Poissonian with $F_{3}<1$. We have found that the most
suppressed photocount noise, given by $F_{3}\approx 5/6$, is
obtained for the balanced initial intensities $r^2_{1}=r_{2}^2$
of the sub-frequency modes and the sum-frequency intensity equal
to $r^2_{3}=r^{2}_{1}/2$ as determined from condition (\ref{N06})
for the no-energy-transfer regime. Scaling laws and polynomial
fits for the maximum sub-Poissonian behavior have been found for
different processes and initial conditions.  We have compared in
detail the non-degenerate and degenerate conversions on time
scales short and long compared to the revival times. We have
observed that the non-degenerate three-wave mixing, contrary to
the degenerate conversion, exhibits stabilization of the
suppressed photon-number noise even on the revival time scale.

\section*{Acknowledgments}

We thank Prof. Jan Pe\v{r}ina and Dr. Ond\v{r}ej Haderka for
helpful discussions. AM is deeply indebted to Prof. Nobuyuki Imoto
for his hospitality and stimulating research at SOKEN. JB was
supported by the Czech Ministry of Education (Grants No. LN00A015
and CEZ J14/98) and the Grant Agency of Czech Republic
(202/00/0142).


{\setlength{\fboxsep}{10pt}
\begin{center}
\framebox{\parbox{0.75\columnwidth}{%
\begin{center}
to be published in\\
{\em J. Opt. B: Quantum Semicl. Opt.}\\
vol. 3 (2001)
\end{center}}}
\end{center}

\end{document}